\documentclass{PoS}

\newcommand{\be}{\begin{equation}}
\newcommand{\ee}{\end{equation}}
\newcommand{\bea}{\begin{eqnarray}}
\newcommand{\eea}{\end{eqnarray}}

\title{Gauged $U(1)$ Clockwork}

\ShortTitle{Gauged $U(1)$ Clockwork}

\author{\speaker{Hyun Min Lee}\thanks{The work is supported in part by Basic Science Research Program through the National Research Foundation of Korea (NRF) funded by the Ministry of Education, Science and Technology (NRF-2016R1A2B4008759). }\\
        Department of Physics, Chung-Ang University, Seoul 06974, Korea.\\
        E-mail: \email{hminlee@cau.ac.kr}}

\abstract{We review the gauged $U(1)$ clockwork models as portals for new physics to obtain effectively hierarchical couplings of the Standard Model (SM) particles from genuine couplings of order one.  Nearest-neighbor mass mixing between multiple $U(1)$ gauge bosons lead to the localization of the zero mode gauge boson, realizing the milicharged gauge boson in the effective theory.  We discuss some of interesting examples utilizing the gauged clockwork beyond the SM.
}

\FullConference{Corfu Summer Institute 2017 'School and Workshops on Elementary Particle Physics and Gravity'\\
		2-28 September 2017\\
		Corfu, Greece}

\begin{document}

\section{Introduction}

Particle physics experiments at intensity and energy frontiers have been complementary direct probes for new physics with light or heavy particles beyond the Standard Model. 
New symmetry for solving the Higgs mass hierarchy problem suggests new particles at weak scale to cancel the quadratic sensitivity to UV physics, but there is no direct hint for new physics yet. However, there are plenty of hints for new physics such as dark matter, strong CP problem, origin of baryon asymmetry, neutrino masses, etc.  Searching in the unknown territory,  minimal approaches for new physics are taken, such as portal models of the following type,
\bea
{\cal L}_{\rm portal}= |H|^2 {\cal O}_{\rm new}, \quad F_{\mu\nu}F^{\prime\mu\nu}, \quad N_R{\cal O}_{\rm new}, \quad a_{\rm ALP} F_{\mu\nu} F^{\mu\nu}, \quad \frac{1}{\Lambda}\, G_{\mu\nu} T^{\mu\nu},\cdots 
\eea
which are portals via Higgs, hypercharge gauge boson, right-handed neutrino, axion-like particle, massive spin-2 particle, etc. 

Recently there have been new interests in the clockwork mechanism \cite{clockwork,gian1}, where hierarchical couplings between the mediators and the SM particles are naturally generated due to the nearest neighbor interactions of the mediators. In this framework, for order one couplings of similar strength to start with, the lightest mass eigenstate of the mediators can have exponentially small couplings to the SM particles due to the localization in the flavor space.  The gravity dual of the clockwork scenarios in extra dimensional models with dilaton background have been discussed \cite{gian1,craig,gian2,choi2,5d2} and has drawn new attention recently.
In this article, we review the basic idea of the clockwork mechanism and discuss the gauged $U(1)$ clockwork in specific and illustrate some interesting examples to resolve some of hierarchy problems in particle physics.

\section{Clockwork mechanism and continuum limit}

Suppose that there are multiple real scalar fields, $\phi_i (i=0,1,2,\cdots,N)$, whose shift symmetries are broken into one shift symmetry, by the following nearest-neighbor interactions \cite{clockwork,gian1},
\bea
{\cal L}_{SCW}= \frac{1}{2} \sum_{j=0}^N \partial_\mu\phi_i \partial^\mu\phi_j -\frac{m^2}{2} \sum_{j=0}^{N-1} (\phi_i- q \phi_{j+1})^2.  \label{sclock}
\eea
Then, the mass matrix for scalar fields is given by
\bea
M^2_\phi= m^2 \left(\begin{array}{cccccc} 1 & -q & 0 &  \cdots &  & 0 \\  -q & 1+q^2 & -q  & \cdots & & 0 \\ 0 & -q  & 1+q^2 & \cdots &  & 0  \\   \vdots & \vdots & \vdots & \ddots & & \vdots  \\  & & &  & 1+q^2 & -q \\ 0 & 0 &  0 & \cdots & -q & q^2 \end{array}  \right).
\eea
The Lagrangian (\ref{sclock}) respects the unbroken shift symmetry, under $\phi_j\rightarrow \phi_j+\frac{c}{q^j}$.  Then, solving the equations for $\phi_j$ leads to $\phi_N=\frac{1}{q}\,\phi_{N-1}=\cdots=\frac{1}{q^N}\,\phi_0$, corresponding to a sequential decrease of field values in going from site 0 toward site $N$  \footnote{We can introduce the nearest-neighbor interactions by $(q\phi_i - \phi_{j+1})^2$ instead. In this case, the field values are related by $\phi_0=\frac{1}{q}\,\phi_{1}=\cdots=\frac{1}{q^N}\,\phi_N$ in the minimum of the potential so they increase in going from site 0 toward site $N$ }. This is the so called clockwork mechanism and the fields in the flavor space are clock gears.  
Indeed, the zero mode has a profile, ${\tilde\phi}_0\sim \sum_j \phi_j /q^j$, localized toward the site at $j=0$, so it has position-dependent couplings to external fields, in particular, suppressed couplings to external fields containing interactions only to $\phi_0$ by ${\cal L}_{\rm int}=\phi_0 {\cal O}_{\rm ext}$. On the other hand, the massive modes of the clockwork, ${\tilde \phi}_k (k\neq 0)$ with nonzero mass $M_k$, are given by the rest of linear combinations of the clock gears, with a mass gap $m$ from the zero mode and the squeezed mass spectrum, $\delta M_k/M_k\sim 1/N$, for a large $N$ \cite{gian1}. 

One can take the continuum limit of the clockwork by introducing a five-dimensional scalar field, $\phi(x,y)=\phi_j(x)$, with $y=ja$ where $a$ is the lattice distance. In the limit of $a\rightarrow 0$ and $N\rightarrow \infty$, we keep $\pi R\equiv N\,a$ finite. Then, additionally taking $m\rightarrow\infty$ with $m\,a\rightarrow 1$ and $q\rightarrow 1$, the clockwork Lagrangian (\ref{sclock}) with interactions to external fields,  ${\cal L}_{\rm int}=\phi_0 {\cal O}_{\rm ext}$, becomes \cite{gian1}
\bea
{\cal L}_{5D,SCW}= \int^{\pi R}_0 dy \left(\frac{1}{2}(\partial_\mu\phi)^2 -\frac{1}{2}(\partial_y\phi+k\phi)^2 \right) + \int^{\pi R}_0 dy \delta(y-\pi R) \phi {\cal O}_{\rm ext}  
\eea
where $k\equiv (q-1)/(qa) $ and $q^N=e^{k \pi R}$. 
Then, the continuum clockwork can be described by a bulk scalar field $\phi$ with bulk and mass terms proportional to $k$ \cite{u1cw}, so the zero mode of $\phi$ has an exponential profile, $\phi_0\sim e^{-ky}$, such that it has suppressed couplings to the external fields.   
With the field redefinition by $\chi= e^{ky} \phi$ and the dilaton background $S=-2ky$, the above Lagrangian becomes \cite{craig,gian2}
\bea
{\cal L}_{5D,SCW}= \int^{\pi R}_0 dy\, e^S \Big[\frac{1}{2}\partial_M\chi \partial^M\chi \Big]+\int^{\pi R}_0 dy \delta(y-\pi R)  e^{\frac{1}{2}S} \chi {\cal O}_{\rm ext}.
\eea
Thus, one bulk scalar field $\chi$ does not have induced mass terms but it has position-dependent kinetic terms. In this picture, the zero mode of the bulk scalar field has a constant bulk profile so it has suppressed couples to the external fields via the dilaton factor.  

We remark the five-dimensional realization for the continuum clockwork.
The 5D gravity with a dilaton $S$ \cite{5d1,gian1} was considered as in the following,
\bea
S&=&\int d^5x \sqrt{-g}\, \frac{M^3_5}{2} \, e^S \Big( R+(\partial_M S)^2 + 4k^2 \Big) \nonumber \\
&&- \int d^5x \sqrt{-g} \,e^S \Big(\frac{\delta(y)}{\sqrt{g_{55}}}\,\Lambda_0+\frac{\delta(y-\pi R)}{\sqrt{g_{55}}}\,\Lambda_\pi \Big)
\eea
where $M_5$ is the 5D Planck mass, $k$ is the 5D curvature and $\Lambda_{0,\pi}$ are the brane tensions.  Then, the scale symmetry with $S\rightarrow S+c$ and $g_{MN}\rightarrow e^{-2c/3}g_{MN} $ is broken explicitly by a nonzero $k^2$ as well as brane tensions. 
With a Weyl rescaling of the metric, $g_{MN}\rightarrow e^{2S/3} g_{MN}$, the above action becomes
in Einstein frame
\bea
S&=&\int d^5x \sqrt{-g}\, \frac{M^3_5}{2} \, \Big( R-\frac{1}{3}(\partial_M S)^2 + 4k^2\, e^{-2S/3}\Big) \nonumber \\
&&- \int d^5x \sqrt{-g} \,e^{-S/3} \Big(\frac{\delta(y)}{\sqrt{g_{55}}}\,\Lambda_0+\frac{\delta(y-\pi R)}{\sqrt{g_{55}}}\,\Lambda_\pi \Big)
\eea
Thus, the brane tensions couple to the dilaton in Einstein frame, stabilizing the radius of the extra dimension.  A 5D dilaton background with 4D flat space was obtained in Einstein frame for the above action, as follows \cite{5d10,5d1,gian1,5d2},
\bea
ds^2 = e^{\frac{4}{3}k|y|} (\eta_{\mu\nu}dx^\mu dx^\nu+dy^2); \quad S=2k|y|,
\eea
with the tuning relation between brane tensions,
\be
\Lambda_\pi=-\Lambda_0=4k M^3_5. 
\ee
In this model, the 4D effective Planck mass is given by
\bea
M^2_P=M^3_5\int^{\pi R}_{-\pi R} dy \sqrt{-g} \,e^{-\frac{4}{3}k|y|}=\frac{M^3_5}{3}\,L_5 \, e^{\frac{1}{3}k\pi R}
\eea
with $L_5\equiv \int^{\pi R}_{-\pi R} dy\, e^{\frac{4}{3}k|y|}$ being the proper length of the extra dimension. Thus, in the 5D dilaton background with branes, the hierarchy problem between the Planck scale and the weak scale can be solved in combination of the warp factor and the large fifth dimension.
The bulk theory is known to be a holographic dual of Little String Theory \cite{littlestring,5d10}.

\section{Gauged $U(1)$ clockwork}

The clockwork mechanism can be simply applied to the case with gauge bosons for multiple local $U(1)$'s. We break $N+1$ copies of local $U(1)$'s, $U(1)_0\times U(1)_1\times \cdots\times U(1)_N$, down to one $U(1)$ by introducing gauge boson masses via Stueckelberg mechanism or Higgs mechanism \cite{gian1,craig,u1cw}, as follows,
\bea
{\cal L}_{GCW}= \sum_{j=0}^{N-1} \frac{1}{2} m^2 \
\Big(A^j_\mu - q A^{j+1}_\mu+\frac{1}{m}\,\partial_\mu\pi^j\Big)^2 \label{GCW}
\eea
where $\pi_j$ transforms under the gauge transformations as $\pi^jj\rightarrow -m(\alpha^j-q \alpha^{j+1})$ for $A^j_\mu\rightarrow A^j_\mu+\partial_\mu\alpha^j$. Here, we note that the remaining $U(1)$ transforms is defined by $A^j_\mu\rightarrow A^j_\mu+\frac{1}{q^j} \partial_\mu\alpha$ under which $\pi^j$ is invariant. 
In the case with Higgs mechanism, we get $m^2=g^2 f^2$ for nonzero VEVs of $\phi_j$ carrying charge $(1,-q)$ under $U(1)_j\times U(1)_{j+1}$, i.e. $\phi_j=\frac{1}{\sqrt{2}}(f+h_j)\, e^{i\pi^j/f}$ \cite{u1cw}.  
The Higgs perturbations of $\phi_j$ turn out to be decoupled in the continuum limit \cite{u1cw}, but they should be kept for the renormalizability of the discrete clockwork.  

Similarly as in the scalar clockwork \cite{gian1}, the zero mode of the gauge clockwork is localized at $j=0$ for $q>1$, given \cite{gian1,u1cw} by
\bea
{\tilde A}^0_\mu=\sum_{j=0}^N a_{j0} A^j_\mu(x)
\eea
where $a_{j0}=N_0/q^j$ with $N_0=\sqrt{(q^2-1)/(q^2-a^{-2N})}$.
The massive modes of the gauge clockwork are also given by
\bea
{\tilde A}^k_\mu(x)=\sum_{j=0}^N a_{jk} A^j_\mu(x),\quad k=1,2,\cdots, N,
\eea
with the mass eigenvalues being
\bea
M^2_k=m^2 \Big( 1+q^2-2q \cos\frac{k\pi}{N+1}\Big)\equiv m^2\lambda_k, \label{mass}
\eea
and the wave functions being 
\bea
a_{jk}= N_k \left[q\sin\Big(\frac{jk\pi}{N+1} \Big)-\sin\Big(\frac{(j+1)k\pi}{N+1} \Big) \right], \quad N_k=\sqrt{\frac{2}{(N+1)\lambda_k}}.  \label{wf}
\eea
The interacting gauge fields are are invertible to get 
\be
A^j_\mu(x)=\sum^N_{j=0} a_{jk} {\tilde A}^j_\mu(x), \quad j=0,1,2,\cdots,N.  \label{flavor}
\ee
This relation is important to identify the gauged clockwork couplings to the external fields localized at a site. 

We can also take the continuum limit of the gauged clockwork by taking $a\rightarrow 0$ and $f\rightarrow\infty$ and $q\rightarrow 1$ with $agfq=1$. Then, with a five-dimensional gauge field, $A_\mu(x,y)\equiv A^j_\mu(x)$ and $A_y(x,y)\equiv \pi^j(x)$, the gauged clockwork Lagrangian becomes \cite{u1cw}
\bea
{\cal L}_{5D,GCW}= \int^{\pi R}_0 dy \left[-\frac{1}{4} F_{\mu\nu} F^{\mu\nu}-\frac{1}{2} (\partial_y A_\mu-\partial_\mu A_y+k A_\mu)^2 \right]
\eea
with $k\equiv (q-1)/(qa)$ again. The resulting Lagrangian contains bulk and brane mass terms proportional to $k$ as in the scalar clockwork, leading to the localization of the zero mode of the gauged clockwork \cite{u1cw} such as
\be
\psi_0(y)=N_0\,  e^{-k|y|}, \quad  N_0=\sqrt{\frac{k\pi R}{1-e^{-2k\pi R}}} 
\ee
in the covering space $S^1$ with $-\pi R<y<\pi R$ where $-\pi R<y<0$ is identified with $0<y<\pi R$ by a $Z_2$ symmetry.
Furthermore, the wave functions for massive modes are 
\be
\psi_n(y)=N_n\Big(\cos\frac{ny}{R}-\frac{kR}{n} \sin\frac{n|y|}{R} \Big), \quad N_n=\frac{n}{m_n R},
\ee
with mass eigenvalues being
\bea
m^2_n=k^2 + \frac{n^2}{R^2}.
\eea
The results can be comparable to the case with discrete clockwork in eqs.~(\ref{mass}) and (\ref{wf}).
For $kR\gg 1$, the warp factor of the zero mode, $e^{-y\pi R}$, gets small, so the spectrum of massive modes is squeezed at the scale of $k$.  

With the field redefinition by $B_M=e^{ky} A_M$, the above Lagrangian can be rewritten as
\bea
{\cal L}_{5D,GCW}= \int^{\pi R}_0 dy\,  e^S \left[-\frac{1}{4} F_{MN} F^{MN}\right]
\eea 
with $F_{MN}=\partial_M B_N -\partial_N B_M$. 
Thus, the continuum gauged clockwork is equivalent to a five-dimensional massless $U(1)$ theory with the dilaton background. 

We consider the gauged clockwork couplings to external fields with nonzero $U(1)$ charges.
General couplings to a fermion $\psi$ localized at site $l$ are given \cite{u1cw} by
\bea
{\cal L}_{\rm fermion}= -g{\bar\psi} \gamma^\mu (v_\psi+a_\psi \gamma^5)\psi  A^l_\mu,
\eea
where $v_\psi, a_\psi$ are vectorial and axial couplings, respectively, 
while those to a scalar $\phi$ localized at site $p$ are given by
\bea
{\cal L}_{\rm scalar}=-ig(\phi^* \partial^\mu \phi-\phi \partial^\mu \phi^*) A^p_\mu + g^2 (A^p_\mu)^2 |\phi|^2. 
\eea
Then, from eq.~(\ref{flavor}), the gauged clockwork couplings to fermions or scalars can be easily identified. In particular, even if fermions or scalars carry order one charge under their local $U(1)$, the effective couplings to the zero mode of the gauged clockwork can be exponentially suppressed  as $g_{\rm eff}= N_0 g/q^{l,p}$, realizing the milicharged $U(1)$.   

Remarkably, when a dark Higgs $\phi$ at site $p$ breaks the remaining $U(1)$ of the gauge clockwork by $\langle\phi\rangle=\frac{1}{\sqrt{2}} v_\phi\ll f$, the mass of the zero mode can be naturally suppressed by $M^2_0=\frac{g^2 N^2_0}{q^{2p}} v^2_\phi\ll v^2_\phi$ \cite{u1cw}. 
From the point view of the continuum clockwork, a localized Higgs mechanism with nonzero VEV $v_{\rm IR}$ on the IR brane at $y=\pi R$ leads to the gauge boson mass which is much smaller than the IR scale by the warp factor as $M_0= e^{-k\pi R}\, g v_{\rm IR}\ll v_{\rm IR}$ \cite{u1cw}.

\section{Examples of gauged $U(1)$ clockwork}

Weakly Interacting Massive Particles (WIMPs) have been the main paradigm for dark matter, having weak-scale mass and neutrino-like interactions to the SM particles. But, there has been no conclusive hint for WIMP dark matter yet and the bounds from direct detection experiments are getting more stringent. On the other hand, orbital velocity curves of galaxies hint at a core dark matter profile, unlike the cuspy profile obtained from the N-body simulation with WIMP dark matter, causing the cusp-core problem. Thus, together with too-big-to-fail problem, there are small-scale problems at galaxy scales. A large self-scattering cross section for dark matter about $\sigma_{\rm self}/m_{\rm DM}=0.1-10\,{\rm cm}^2/g$  would be required to solve the small-scale problems \cite{smallscale}.

Various applications of the gauged clockwork have been proposed in Ref.~\cite{u1cw}, such as self-interacting dark matter with suppressed couplings to the SM particles, flavor-dependent $U(1)'$ interactions for explaining the $B$-meson anomalies in semi-leptonic decays at LHCb \cite{ligong}, and suppressed $D$-term supersymmetry breaking due to gauge kinetic mixings between multiple $U(1)$ gauge multiplets, etc.

Among those concrete models for the gauged clockwork, we consider the gauged clockwork as a mediator between dark matter and the SM particles to explain the smallness of interactions of dark matter to the SM particles while allowing for large self-interactions of dark matter, due to the localization of the light zero mode of a mediator gauge boson towards dark matter. 
There are anomaly-free $U(1)'$ without extra charged fermions  beyond the SM. For instance, $U(1)_{B-L}$ and $U(1)_{L_i-L_j}$ with $i,j=1,2,3$ and $i\neq j$ are such examples \cite{u1cw}. The LHC searches for extra gauge bosons such as dimuon searches, however, have put stringent constraints on the $Z'$ mass for $U(1)'$ with SM hypercharge-like gauge coupling  to be heavier than  multi-${\rm TeV}$.  Thus, in this case, thermal dark matter mediated by one of those gauge interactions would be heavy near the half the gauge boson mass and have very tiny self-interactions. 

In the framework of the gauged clockwork, we introduce general $U(1)'$ interactions to a Dirac fermion dark matter $\chi$ at site 0 and the SM fermions $f$ at site $N$ \cite{u1cw}, as follows,
\bea
{\cal L}_{Z'}= -g_{Z'} {\bar \chi} \gamma^\mu (v_\chi + a_\chi \gamma^5) \chi\,Z^{\prime 0}_\mu  -g_{Z'} {\bar \chi} \gamma^\mu (v_f + a_f \gamma^5) \chi\,Z^{\prime N}_\mu \label{dmmed}
\eea
where $v_{\chi,f}$ and $a_{\chi,f}$ are vectorial and axial couplings for dark matter and SM fermions, respectively. 
Then, expanding the gears of the gauged clockwork in terms of mass eigenstates ${\tilde Z}^k_\mu$ as in eq.~(\ref{flavor}),
\be
Z^{\prime j}_\mu(x)=\sum^N_{j=0} a_{jk} {\tilde Z}^j_\mu(x), \quad j=0,1,2,\cdots,N,
\ee
eq.~(\ref{dmmed}) becomes
\bea
{\cal L}_{Z'}&=& -g_{Z'} {\bar \chi} \gamma^\mu (v_\chi + a_\chi \gamma^5) \chi\,\left( N_0 {\tilde Z}^{ 0}_\mu+\sum_{k=1}^N a_{0k} {\tilde Z}^k_\mu(x)\right)  \nonumber \\
&&  -g_{Z'} {\bar f} \gamma^\mu (v_f + a_f \gamma^5) f\left( \frac{N_0}{q^N} {\tilde Z}^{ 0}_\mu+\sum_{k=1}^N a_{Nk} {\tilde Z}^k_\mu(x)\right).
\eea
As a consequence, for $m_\chi\ll M_0\ll M_k$ with $k\neq 0$, we can integrate out the gauged clockwork to obtain the effective interactions for dark matter,
\bea
{\cal L}_{\rm eff}&=& -\frac{N^2_0 g^2_{Z'}}{2M^2_0} \Big({\bar\chi} \gamma^\mu (v_\chi + a_\chi \gamma^5) \chi\Big)^2 -\frac{N^2_0 g^2_{Z'}}{2 q^{2N}M^2_0} \Big({\bar f} \gamma^\mu (v_f + a_f \gamma^5) f\Big)^2 \nonumber \\
&&-\frac{N^2_0 g^2_{Z'}}{ q^N M^2_0} \Big({\bar\chi} \gamma^\mu (v_\chi + a_\chi \gamma^5) \chi\Big)\Big({\bar f} \gamma^\mu (v_f + a_f \gamma^5)f\Big).  \label{effective}
\eea
As a consequence, the effective self-interactions of dark matter can be large due to the light zero mode of the gauged clockwork, while the interactions between dark matter and SM fermions are suppressed for $q^N\gg 1$ due to a strong localization of the zero mode towards dark matter. 
Therefore, a self-interacting dark matter can be realized with small couplings (or effective milicharges) to the SM particles in the gauged clockwork models \cite{u1cw}.

For a light zero mode of the gauged clockwork, the above effective Lagrangian is not valid and we need to include the light zero mode explicitly in the calculations of dark relic density, direct detection  and self-scattering cross section, etc \cite{u1cw,z3dm}. Dark photon searches at LHCb, Belle II and SHiP experiments \cite{intensity} could probe the zero mode of the gauged clockwork in relation to the self-interacting dark matter.

Furthermore, there are additional effective interactions induced due to massive states of the gauged clockwork, but they are negligible for $M_k\gg q^{N/2} M_0$ for $k\neq 0$. Otherwise, the sum of those massive states contribute sizably to the effective interactions \cite{u1cw,hong}. 
In this case, collider searches for massive states with squeezed spectrum \cite{5d2} would be complementary for testing the gauged clockwork.

\section{Conclusions}

We have given a review on clockwork scenarios where multiple global or local symmetries are broken into a single symmetry due to nearest-neighbor interactions between multiple fields. We discussed the gauged clockwork mechanism in more detail and showed that the localization of the zero mode in the gauged clockwork can explain the hierarchy of couplings in external fields for genuine couplings of order one. Various interesting applications of the clockwork scenarios  are suggested in concrete models with extra $U(1)$ interactions.

\end{document}